\documentclass{aastex62}
\usepackage{mkfig}

\begin{document}

\title{Morphological Reconstruction of a Small Transient Observed
  by {\em Parker Solar Probe} on 2018~November~5}

\author{Brian E. Wood, Phillip Hess, Russell A. Howard,
  Guillermo Stenborg, Yi-Ming Wang}
\affil{Naval Research Laboratory, Space Science Division,
  Washington, DC 20375, USA; brian.wood@nrl.navy.mil}


\begin{abstract}

     On 2018 November 5, about 24 hours before the first close
perihelion passage of {\em Parker Solar Probe} ({\em PSP}), a coronal
mass ejection (CME) entered the field of view of the inner detector of
the Wide-field Imager for Solar PRobe (WISPR) instrument onboard {\em PSP},
with the northward component of its trajectory carrying the leading
edge of the CME off the top edge of the detector about four hours
after its first appearance.  We connect this event to a very small
jet-like transient observed from 1~au by coronagraphs on both the
{\em SOlar and Heliospheric Observatory} ({\em SOHO}) and the A
component of the {\em Solar TErrestrial RElations Observatory}
mission ({\em STEREO-A}).  This allows us to make the first
three-dimensional reconstruction of a CME structure considering
both observations made very close to the Sun and images from two
observatories at 1~au.  The CME may be small and
jet-like as viewed from 1~au, but the close-in vantage point of
{\em PSP}/WISPR demonstrates that it is not intrinsically
jet-like, but instead has a structure consistent with a flux rope
morphology.  Based on its appearance in the {\em SOHO} and
{\em STEREO-A} images, the event belongs in the ``streamer
blob'' class of transients, but its kinematic behavior is very
unusual, with a more impulsive acceleration than previously studied
blobs.

\end{abstract}


\section{Introduction}

     Most studies of coronal mass ejections (CMEs) naturally focus on
the biggest, fastest, and brightest events, which are also the ones
that are the most geoeffective when they happen to hit Earth.  However,
CMEs come in a wide range of sizes and speeds, with smaller, slower
events being more numerous than the large, fast ones
\citep{sy04,av17,bew17}.
Due to their greater frequency, the smaller transients may collectively
account for a significant fraction of what is generally regarded as
the quiescent slow solar wind \citep{ekjk09,mj14}.

     For small transients observed in white-light (WL) images, it can
be ambiguous whether ``coronal mass ejection'' is the best descriptive
moniker.  For example, using observations from the Large Angle
Spectrometric COronagraph (LASCO) instrument
on board the {\em SOlar and Heliospheric Observatory} ({\em SOHO}),
\citet{nrs97} identified a class of small jet-like transients
emanating from the tops of helmet streamers, which could generically
be termed ``streamer disconnection events,'' but have been more
informally called ``streamer blobs'' \citep{ymw98}.  These
events have slow acceleration profiles that are believed to simply
track the ambient slow solar wind into which they are released \citep{ihc18}.
Likewise, CMEs that have no associated solar surface activity
have similarly slow acceleration profiles, and though some of
these events can be large and bright, some are as small and faint as
the streamer blobs \citep{bew17}.  The distinction between
streamer blobs and the slow CMEs is also blurred by
morphological similarity.  The stereoscopic imaging capabilities
provided by the {\em Solar TErrestrial RElations Observatory}
({\em STEREO}) mission have provided evidence that streamer blobs
have a flux rope (FR) structure \citep{nrs09,apr11},
consistent with the most favored interpretation of CME
morphology \citep{jc97,vb98,at09,av13,bew17}.

     Even further along the spectrum from large, obvious transients
toward smaller density enhancements within the solar wind
are the periodic density structures identified in coronagraphic and
heliospheric images \citep{nmv10,nmv15}.
\citet{ced18} find that such compact solar wind
structures possess a continuum of sizes, with the streamer blobs at
the large-scale end.  Some of these structures might be associated
with reconnection among streamer loops, others with
interchange reconnection between closed loops and open flux
at coronal hole boundaries, as suggested by recent numerical
modeling \citep{akh17,akh18}.

     The launch of {\em Parker Solar Probe} ({\em PSP}) on
2018~August~12 provides an opportunity to study small solar wind
transients from a vantage point closer to the Sun, allowing
a more detailed inspection of their structure.  During each close
perihelion passage of PSP, the plasma instruments can
detect such transients in~situ, and the Wide-field Imager for Solar
PRobe (WISPR) instrument can provide close-up images of these
events.  The first perihelion passage
occurred on 2018~November~6, with {\em PSP} reaching 35.4~R$_{\odot}$
from Sun-center.  We here report on a small CME observed by WISPR a
day earlier, which is also observed by coronagraphs on both {\em SOHO}
and {\em STEREO-A}.  The resulting stereoscopic imaging allows
us for the first time to study CME morphology considering both
multiple vantage points at 1~au and a viewpoint very close to the Sun.

\section{Observations}

\begin{figure}[t]
\plotfiddle{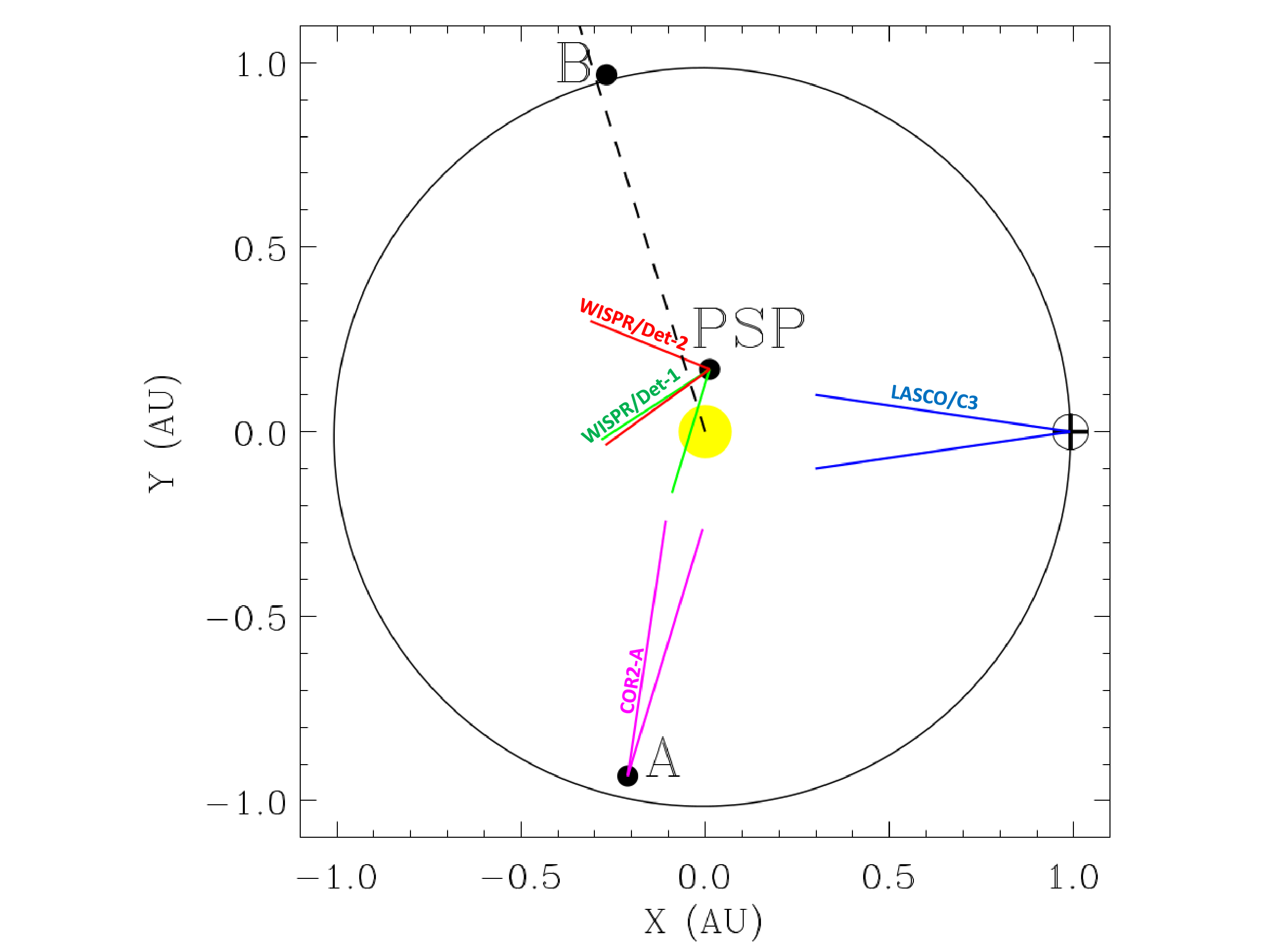}{3.6in}{0}{55}{55}{-220}{-5}
\caption{The positions of Earth, {\em PSP}, {\em STEREO-A}, and
  {\em STEREO-B} in the ecliptic plane on 2018~November~5 (in
  HEE coordinates).  The blue lines indicate
  the FOV of the LASCO/C3 coronagraph on {\em SOHO},
  near Earth.  The purple lines indicate the FOV of COR2-A on {\em
  STEREO-A}, and the green and red lines are the FOV's of WISPR's
  Detector~1 and Detector~2, respectively.  The dashed line indicates
  the central trajectory of the small CME observed by LASCO,
  COR2-A, and WISPR on November~4-5.}
\end{figure}
     The first perihelion passage of {\em PSP} occurred at UT 03:28 on
2018~November~6, at 35.4~R$_{\odot}$ from Sun-center.  About 24 hours
earlier, the WISPR imager on {\em PSP} observed the small CME that is
the subject of this article, which was also detected by {\em
SOHO}/LASCO and {\em STEREO-A}.  Understanding these observations
requires knowledge of the viewing geometry involved, which is shown in
Figure~1.  A heliocentric Earth ecliptic (HEE) coordinate system is
used, with the x-axis pointed from the Sun toward Earth, and the
z-axis pointed from the Sun toward ecliptic north.  The {\em SOHO}
spacecraft is located near Earth, specifically at the L1 Lagrangian
point.  The LASCO instrument on {\em SOHO} includes two coronagraphs,
C2 and C3, observing the WL corona at Sun-center distances in the
plane-of-sky of 1.5--6 R$_{\odot}$ and 3.7--30 R$_{\odot}$,
respectively \citep{geb95}.  The field of view (FOV) of the
latter is shown explicitly in Figure~1.  Also shown is the FOV of the
COR2-A coronagraph on {\em STEREO-A}, covering 2.5--15.6 R$_{\odot}$
\citep{rah08}.

     The WISPR instrument on {\em PSP} possesses two heliospheric
imagers, Detector~1 and Detector~2, which look in the ram direction of
{\em PSP}'s orbit around the Sun, covering elongation angles from
Sun-center of $13^{\circ}-53^{\circ}$ and $50^{\circ}-108^{\circ}$,
respectively \citep{av16}.  These FOVs are also explicitly
shown in Figure~1, although the CME in question never actually enters
the Detector~2 FOV.  Although {\em STEREO-B}'s location is indicated,
observations from it are unfortunately not available, as
{\em STEREO-B} has not been operational since late 2014.

\begin{figure}[t]
\plotfiddle{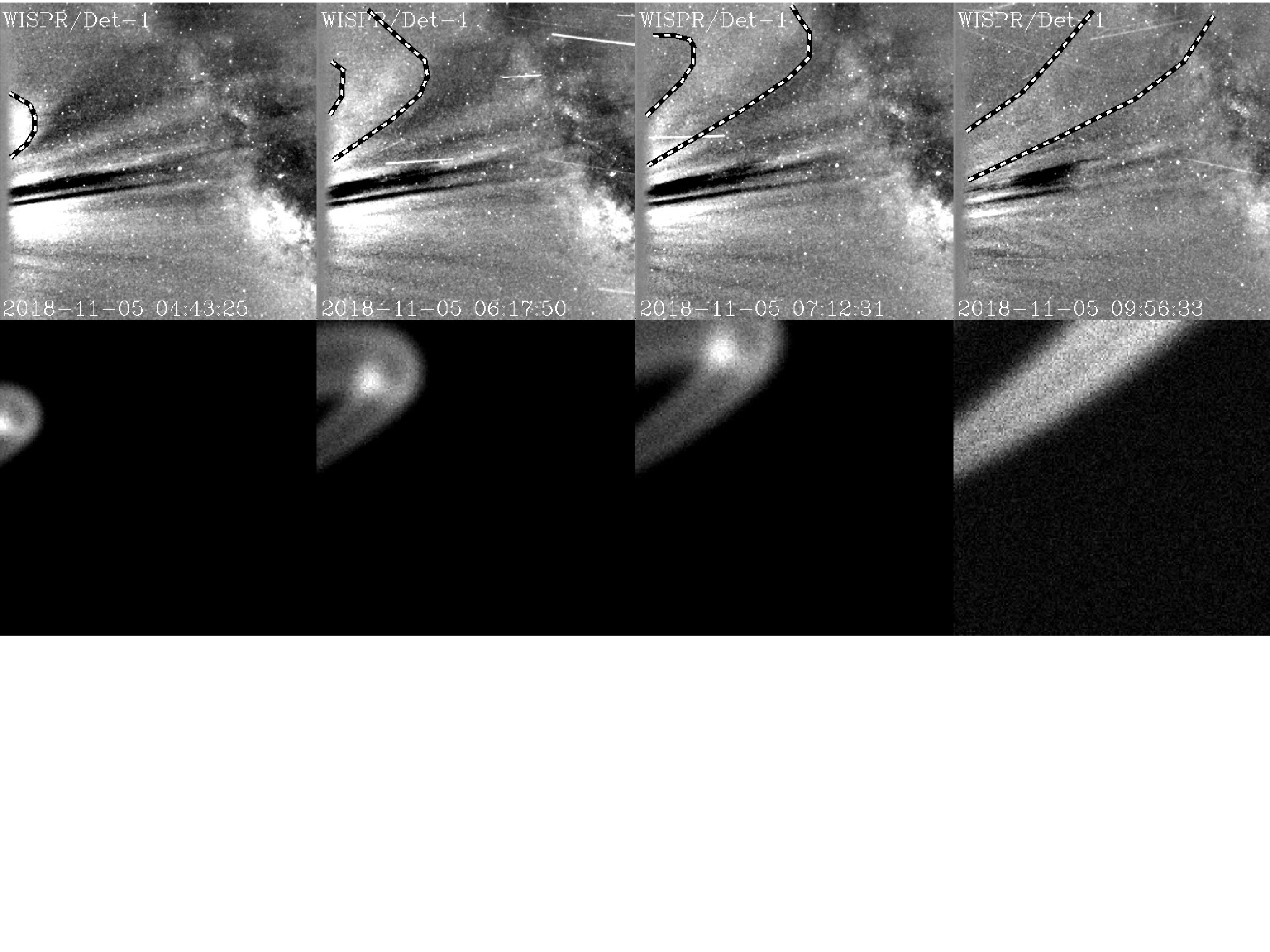}{3.0in}{0}{64}{64}{-230}{-120}
\caption{Images of the 2018~November~5 CME from Detector~1 of the
  WISPR instrument on {\em PSP}.  The images are cropped on the
  bottom and right to focus attention on the region of interest.
  Dotted lines outline the transient.  The Milky Way stretches vertically
  across the right side of the images.  Synthetic images of the event
  are shown below the real images, based on the 3-D reconstruction
  described in Section~3.  A movie version of this figure is
  available online.}
\end{figure}
     Figure~2 shows a sequence of four WISPR Detector~1 images
from UT 4:43 to UT 9:56 on 2018~November~5, with a small CME entering
on the upper left side of the detector and eventually exiting off the
top of the FOV.  Generating such WL images requires first removing the
dust-scattered F-corona contribution, which is performed as part of
the WISPR data processing, analogous to what has been done for
STEREO data \citep[e.g.,][]{gs17}.  This is
somewhat more complicated for WISPR than for WL images from 1~au
observatories, because {\em PSP}'s distance from the Sun is changing
significantly during the observations, and therefore the F-corona
background is time dependent.  Removal of this background yields a WL
image that includes only the K-corona contribution of interest, which
is due to Thomson scattering from electrons.  The K corona images are
dominated by emission from the quiescent streamer structure, but we
wish to focus on the emission from the transient emission.  Thus, we
compute an average K corona image from November~5 and then subtract
this from all the images to emphasize the variable emission.  We also
use median filtering to suppress stellar point sources.  The images in
Figure~2 have been cropped on the bottom and right sides to focus
attention on the location where the CME is observed.

\begin{figure}[t]
\plotfiddle{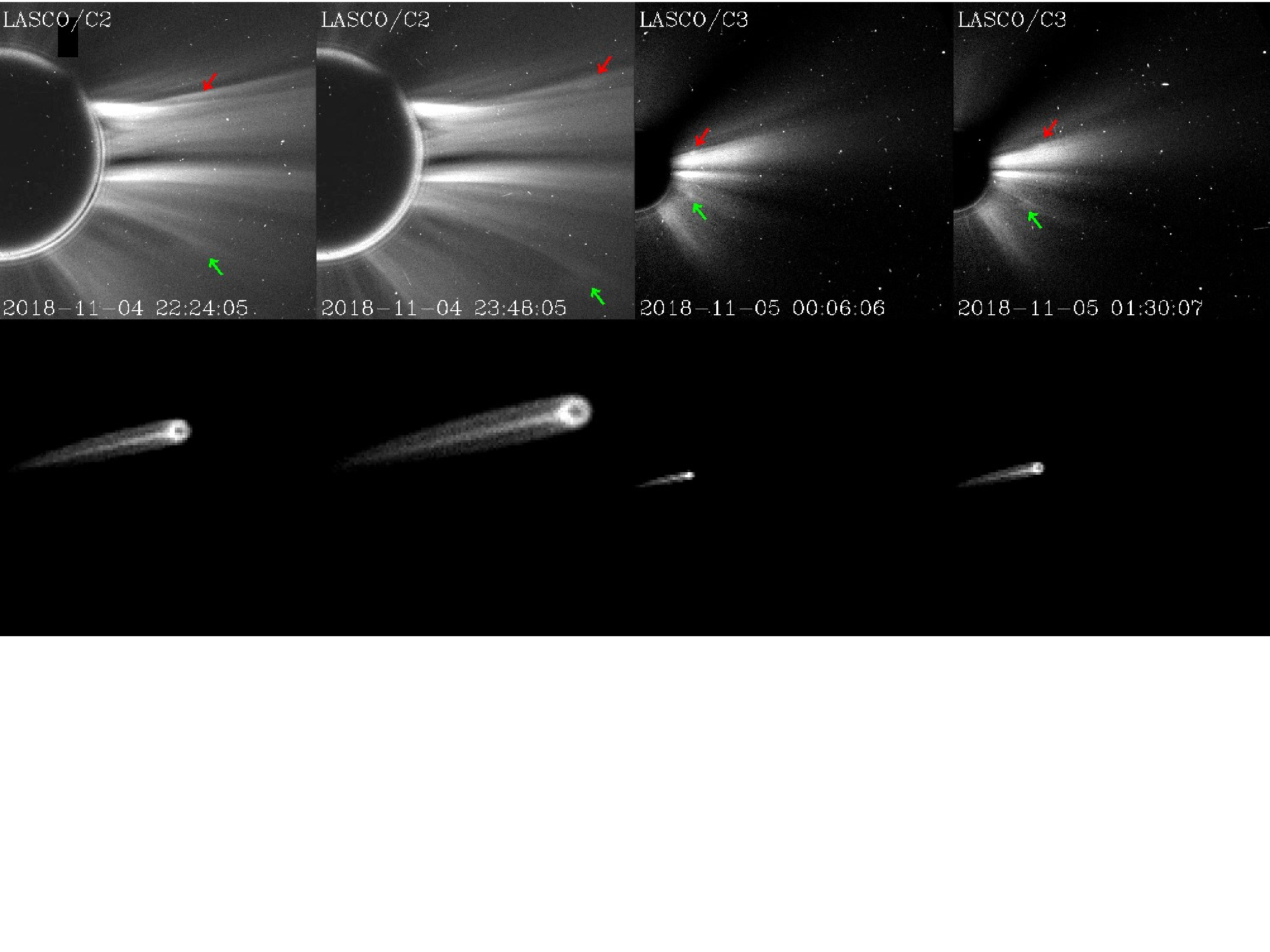}{3.0in}{0}{64}{64}{-230}{-120}
\caption{A sequence of images from the LASCO C2 and C3 coronagraphs
  on board {\em SOHO}, showing two little transients whose positions
  are marked by red and green arrows.  It is the upper event that
  corresponds with the CME observed by WISPR.  Synthetic images of
  this transient are shown below the real images, based on the 3-D
  reconstruction described in Section~3.  A movie version of this
  figure is available online.}
\end{figure}
\begin{figure}[t]
\plotfiddle{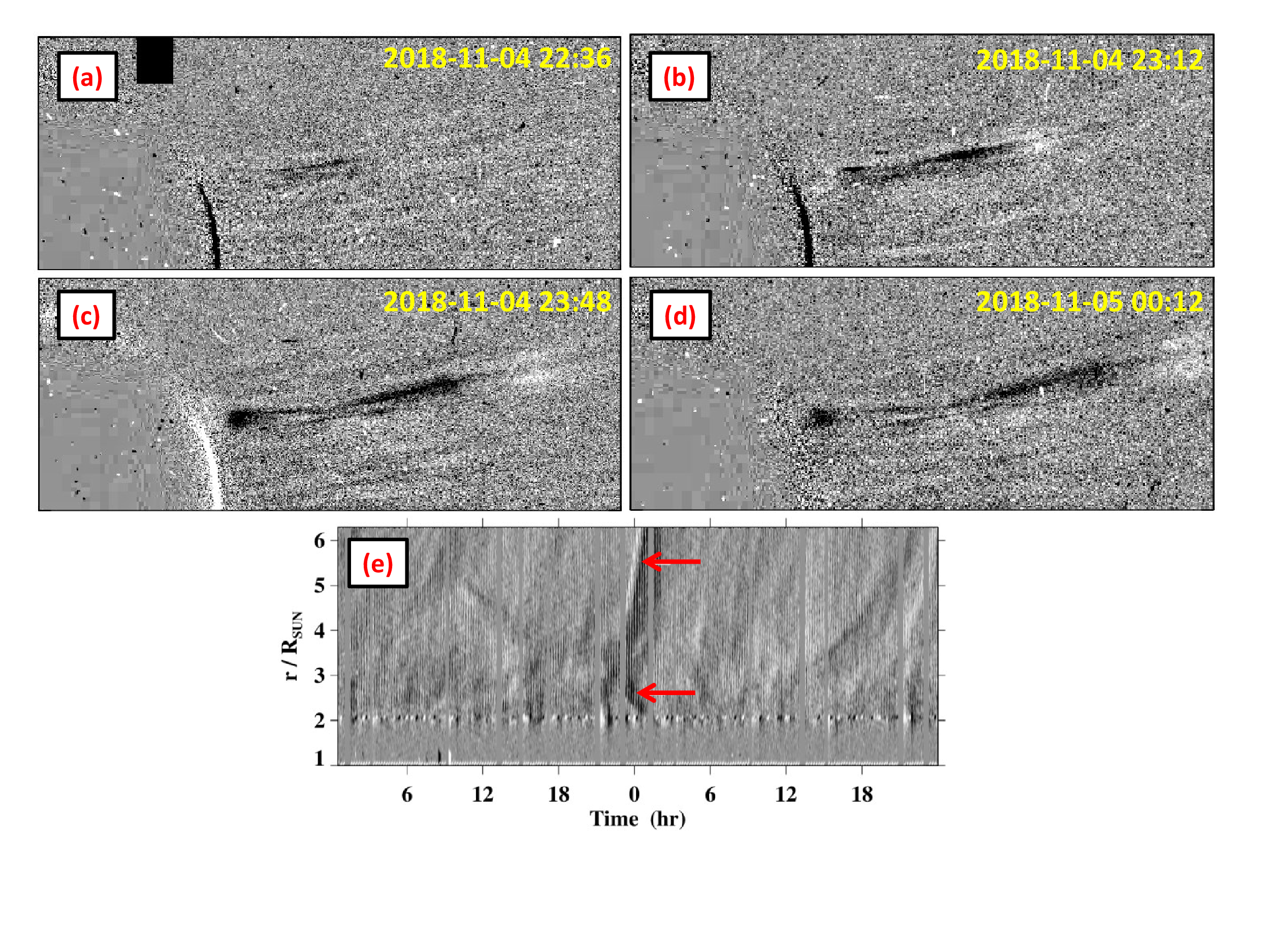}{3.3in}{0}{60}{60}{-215}{-50}
\caption{(a-d) A sequence of four running-difference images from the LASCO C2
  coronagraph, focused on the transient observed by WISPR.
  (e) A height-time stack plot for the position angle of the
  transient, with arrows pointing towards the outflow and below it an
  apparent inflow.}
\end{figure}
     We connect the November~5 CME observed by WISPR with a tiny
transient observed by {\em SOHO}/LASCO on November~4--5.  Figures~3 and 4
show a sequence of LASCO images from this time period.
In Figure~3 the images are shown after subtracting a monthly minimum image,
which preserves the appearance of the quiescent streamer structure, while
in Figure~4, four C2 images are shown in a running difference format,
which makes the small transient easier to see.
There are actually two small transients observed by LASCO, one
to the south marked by a green arrow in Figure~3, and one to the north
marked by a red arrow.  It is the latter that corresponds to the CME
observed by WISPR.  The simultaneity of the two transients
suggests that they are related, both possibly resulting from a minor
adjustment to the large-scale streamer structure on the west side
of the Sun.  In terms of their small size and clear association with
the streamer structure, both events seem consistent with the
``streamer blob'' class of transients \citep{nrs97,ymw98},
although we will show below that the WISPR-observed event
has an impulsive acceleration that is very atypical for streamer blobs.
It is worth noting that despite their small size, both of
these events make it into the online CDAW CME
catalog \citep[][https://cdaw.gsfc.nasa.gov/CME\_list]{sy04}, with
listed start times of UT 22:12 and UT 22:36 on November~4 for the
south and north events, respectively.

     The origin of the event of interest is clearest in Figure~4,
where the outflow begins in the middle of the FOV.  There is
evidence for a pinch-off region, like the events described by
\citet{nrs07}, leading not only to the outflowing
blob but also an apparent downflow below it.  This is clarified in
the height-time stack plot shown in Figure~4(e), where traces of
the running-difference C2 intensities at the position angle of the
event are stacked.  In this plot, the transient is seen as parallel
bright and dark streaks with positive slopes, indicating an outflow.
A dark streak below it with a negative slope indicates the
apparent downflow.

\begin{figure}[t]
\plotfiddle{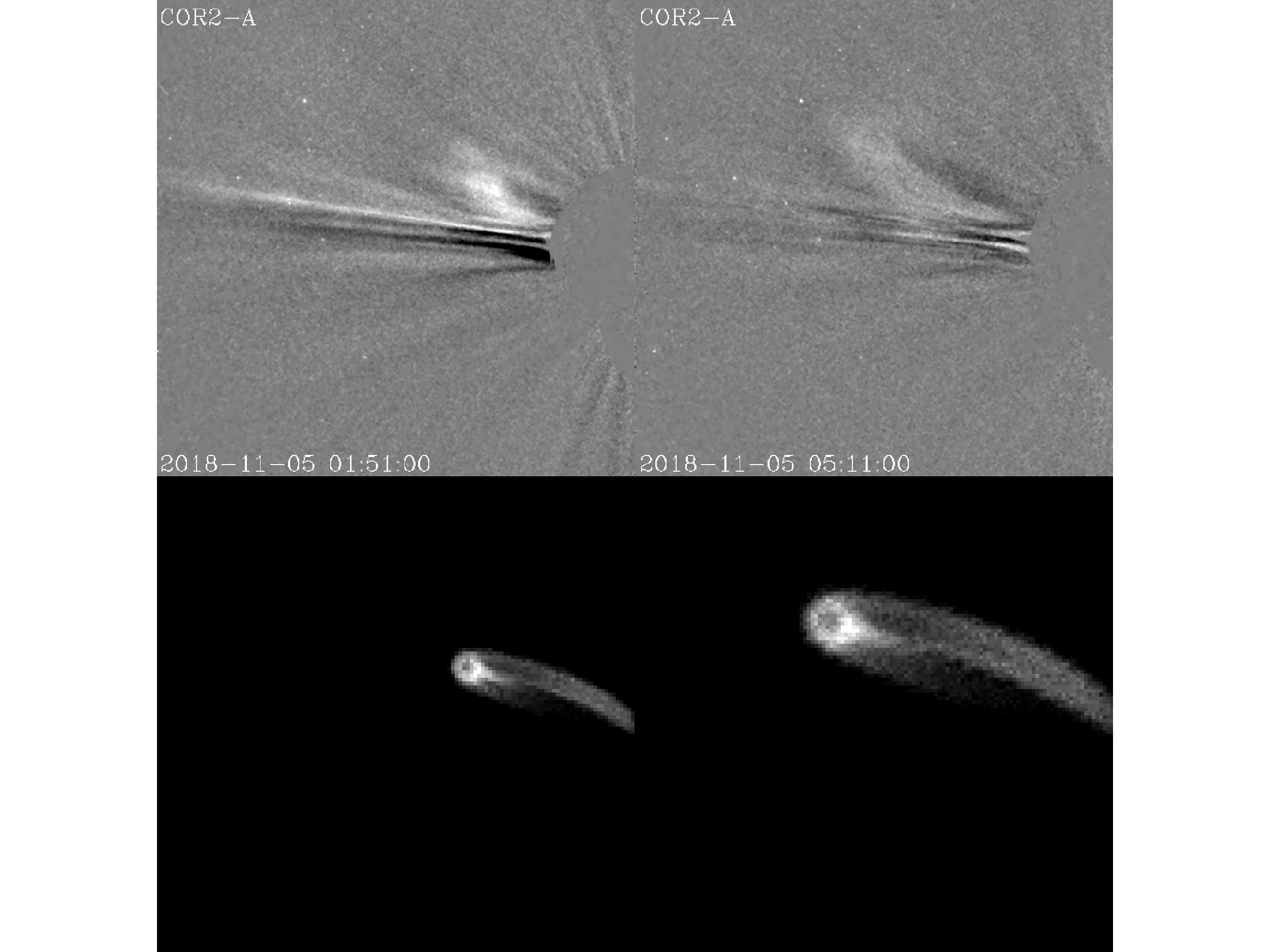}{3.0in}{0}{47}{47}{-170}{-5}
\caption{A sequence of two images of the 2018~November~5 CME
  from the COR2-A coronagraph on board {\em STEREO-A}.  Synthetic
  images of this transient are shown below the real images, based
  on the 3-D reconstruction described in Section~3.  A movie
  version of this figure is available online.}
\end{figure}
     The small WISPR CME is also observed by COR2-A, as shown
in Figure~5.  These images are shown after the subtraction of
an average COR2-A image from November~5, in order to better reveal
the faint transient.  The angular extent of the CME in COR2-A
is somewhat wider than in the LASCO images in Figures~3-4, where
the CME is particularly narrow and jet-like.

     It is not surprising that the CME would look much bigger in
the WISPR images than in LASCO and COR2-A images from 1~au,
considering how much closer {\em PSP} is to the event (see Figure~1).
However, the most important characteristic to note about
the CME's appearance is that unlike in the LASCO
and COR2-A images, the CME in WISPR is not jet-like at all.  Instead,
the transient in WISPR looks very much like an FR structure,
with two legs stretching back towards the Sun.
Its appearance is dominated by one leg that ultimately
stretches through the FOV from east to north, but at early times
it is clear that the CME structure is not linear but bends up and
then backwards, presumably curving into a second leg that is almost
entirely above the FOV.  It is also noteworthy that the lower
leg remains visible long after the leading edge of the CME has
moved out of the FOV, suggesting continued magnetic connectivity
with the Sun.

     Magnetic FRs can be described as tube-shaped structures
permeated by a helical magnetic field, with legs that stretch back
towards the Sun \citep{vb98}.  Evidence that FRs
are at the core of all CMEs comes from both in~situ plasma
measurements \citep{rpl90,igr96}
and WL imaging \citep{jc97,at09,av13,bew17}.  Our observations
provide further support for this interpretation, demonstrating
that small CMEs that look linear and jet-like from 1~au are
revealed to have an FR appearance when viewed up close.  We
now perform a detailed three-dimensional (3-D) reconstruction of
the CME to demonstrate this more explicitly.

\section{Morphological Reconstruction}

     We reconstruct the 3-D structure of the CME assuming an
intrinsic FR shape, using techniques that have been developed to
interpret CME images from the {\em STEREO} spacecraft
\citep{bew09,bew17}.  We refer the reader to this previous work for
details, but in short,
the shape of the inner and outer edges of a 2-D FR are defined
in polar coordinates, and then the two 2-D loops are used to define a
3-D FR shape by assuming a circular cross section for the FR,
bounded by the two loops.  By stretching the FR in the direction
perpendicular to the FR creation plane, an FR can be created with
an arbitrary ellipticity.  The 3-D FR is then rotated into the
desired orientation in an HEE coordinate system.
Adjusting the various quantities involved in the FR creation process
allows experimentation with different shapes and orientations.
For confronting the model FR shape with the actual images of the
CME, mass is first placed onto the
surface of the 3-D FR shape, but not in the interior, and then synthetic
images of the resulting density cube are computed, based on
calculations of the Thomson scattering
within the density cube.  The assessment of the best fit relies on
subjective judgment, with the best-fit parameters discerned through
trial-and-error.  The shape of the CME structure is not assumed to
vary at all, meaning we assume the structure expands in a self-similar
fashion.

\begin{figure}[t]
\plotfiddle{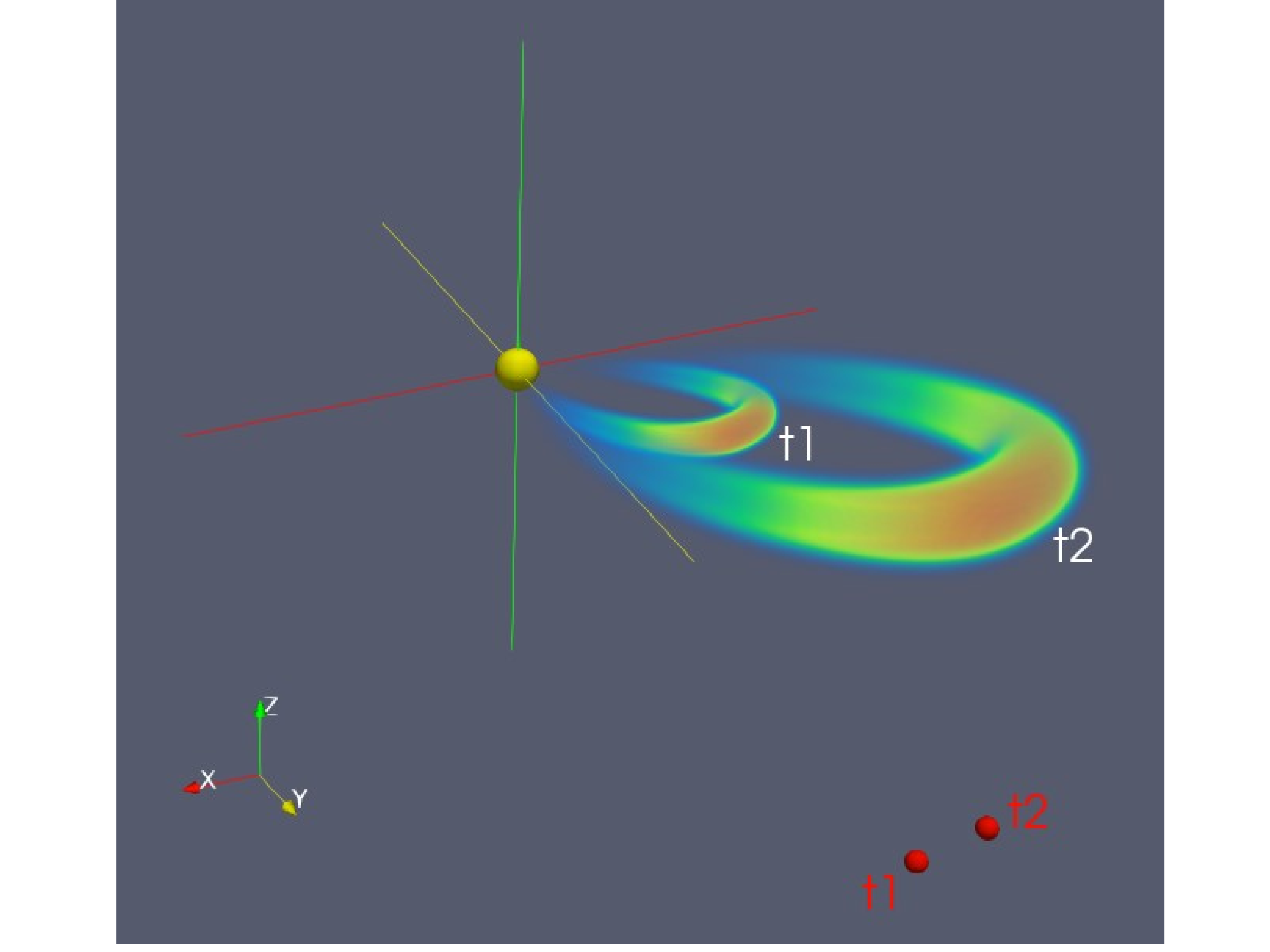}{3.3in}{0}{50}{50}{-180}{-5}
\caption{Reconstructed 3-D FR structure of the CME observed by
  {\em PSP}/WISPR on 2018~November~5, shown in HEE coordinates.  The
  FR is shown at two times, t1 and t2, corresponding to UT 3:48 and
  UT 9:40, respectively.  The red circles indicate the location of
  {\em PSP} at these two times.  The size of the Sun is to scale.}
\end{figure}
     Figure~6 shows our morphological reconstruction of the CME,
in an HEE coordinate system.  The FR is shown at two times,
UT 3:48 and UT 9:40, corresponding to the beginning and end of the
movie version of Figure~2.  The location of {\em PSP} at these two
times is also indicated.  The synthetic images of the CME based on
this reconstruction are shown in Figures~2, 3, and 5, for comparison with the
real images.  The reconstruction successfully reproduces both the
CME's appearance in the WISPR images (Figure~2) and the narrow,
jet-like appearance in the LASCO data (Figure~3).  The reconstructed
FR is almost perfectly edge-on as viewed from {\em SOHO}'s
perspective, explaining why it is so narrow in the LASCO images.

     We consider the synthetic COR2-A images to be an adequate
reproduction of the observations (see Figure~5), but deviations from
the data are larger here.  Specifically, the FR leg that is slightly
higher in the COR2-A images is predicted to be brighter than observed.
This is the left leg in Figure~6, which is the leg that is mostly out
of the WISPR FOV in Figure~2.  We suspect that improved agreement with
the data could be achieved only by introducing asymmetries of some
sort into our FR structure.  Our parametrized FR shape is a symmetric
one, with mass placed onto its surface in a symmetric fashion,
increasing with distance from the Sun as $r^{\beta}$ (see Figure~6),
with the exponent assumed to be $\beta=2.5$.  The easiest way to
improve the COR2-A appearance would be to decrease the mass in the leg
that seems too bright in the synthetic COR2-A images.  This would make
the leg fainter in the COR2-A images, but would not meaningfully
affect the WISPR or LASCO images at all, since that leg is mostly
outside the WISPR FOV, and is superposed on the other, presumably
brighter leg in the LASCO images.

%
\begin{deluxetable}{clc}
\tabletypesize{\scriptsize}
\tablecaption{Flux Rope Parameters}
\tablecolumns{3}
\tablewidth{0pt}
\tablehead{
  \colhead{Parameter} & \colhead{Description} & \colhead{Value}}
\startdata
$\lambda_s$ (deg)& Trajectory longitude             & 107  \\
$\beta_s$ (deg)  & Trajectory latitude              &  13  \\
$\gamma_s$ (deg) & Tilt angle of FR                 &   5  \\
FWHM$_s$ (deg)   & Angular width                    & 43.6 \\
$\Lambda_s$      & Aspect ratio                     & 0.039\\ 
$\eta_s$         & Ellipticity of FR cross section  & 1.0  \\
$\alpha_s$       & Shape parameter for leading edge & 2.5  \\
\enddata
\end{deluxetable}
     Table~1 lists the parameters that define the shape of the
FR, using the variable names from \citet{bew17}.  Briefly,
$\lambda_s$ and $\beta_s$ are the central trajectory in HEE
coordinates, with the $\lambda_s=107^{\circ}$ direction explicitly
indicated in Figure~1.  This longitude is very close to that of
{\em PSP}.  The FR trajectory latitude is only slightly above the
ecliptic, with $\beta_s=13^{\circ}$, but the FR is so thin that this
is high enough for the FR to be entirely above the ecliptic, thereby
missing {\em PSP}.  If $\beta_s$ had instead been less than
$5^{\circ}$, the CME would have likely hit the spacecraft.
Although this FR passed close to the spacecraft, it was not close
enough while in the WISPR FOV to resemble the kinds of FR passages
that \citet{pcl19} modeled in their predictions of what FRs
might look like in WISPR data.

     The $\gamma_s$ parameter indicates the tilt angle of the FR,
with $\gamma_s=0^{\circ}$ corresponding to an E-W orientation parallel
to the ecliptic, and $\gamma_s>0^{\circ}$ indicating an upward tilt of
the west leg.  With $\gamma_s=5^{\circ}$, our FR is close to being
oriented perfectly E-W.  The FWHM$_s$ parameter is the
full-width-at-half-maximum angular width of the FR.  Our
FWHM$_s=43.6^{\circ}$ value places it among the narrowest CMEs in the
\citet{bew17} survey of 28 Earth-directed events.  The aspect
ratio, $\Lambda_s$, indicates the radius of the apex of the FR divided
by the distance of the apex from the Sun, and so is a measure of how
thin the FR is.  The $\Lambda_s=0.039$ value in Table~1 is indicative
of a very thin FR, thinner than any in the \citet{bew17} sample.
Our FR reconstruction scheme allows for the possibility of an
elliptical FR channel, but we see no evidence for ellipticity, so
$\eta_s=1.0$.  Finally, the $\alpha_s$ parameter defines the shape of
the FR leading edge \citep[see][]{bew17}, with higher values
leading to flatter leading edges.  Our rather low $\alpha_s=2.5$ value
yields a leading edge that is better described as curved rather than
flat.

\section{Kinematics}

\begin{figure}[t]
\plotfiddle{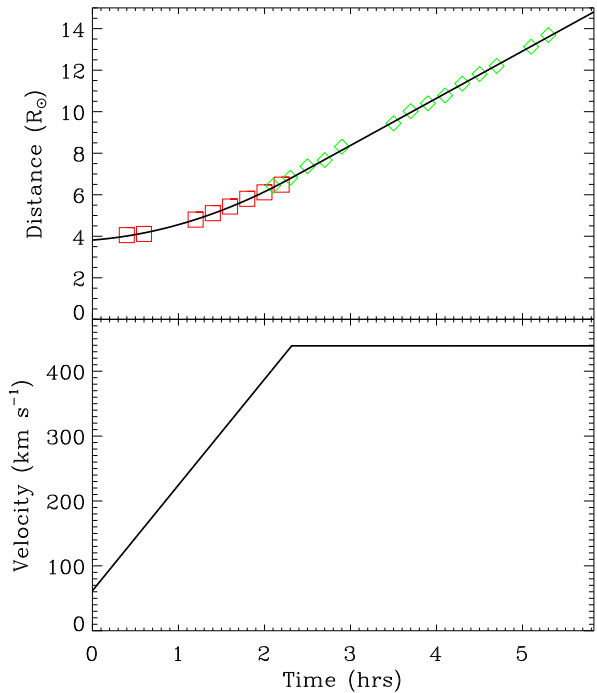}{3.1in}{0}{75}{75}{-220}{-280}
\caption{The top panel shows distance measurements for the leading
  edge of the 2018~November~5 CME as a function of time based on
  images from LASCO/C2 (red squares) and LASCO/C3 (green diamonds).
  The $t=0$ point on the time axis corresponds to UT 22:00 on
  November~4.  These data points are fitted with a simple kinematic
  model assuming a constant acceleration phase followed by a constant
  velocity phase.  The solid line is the best fit, and the bottom
  panel shows the inferred velocity profile.}
\end{figure}
     Computing the synthetic images in Figures~2, 3, and 5 requires not only
the morphological reconstruction shown in Figure~6, but also a
kinematic model to describe how the structure expands with time.  Our
kinematic measurements of the CME are based on the LASCO measurements,
as LASCO has the best vantage point for tracking the leading edge of
the CME.  The top panel of Figure~7 shows distance measurements from
the C2 and C3 coronagraphs, with measured elongation angles converted
to distances assuming the CME trajectory direction found from the
morphological analysis.  In order to infer a velocity profile for the
CME, we fit a simple two-phase kinematic model to the data, assuming a
phase of constant acceleration followed by a phase of constant
velocity.  In the resulting kinematic fit, the CME accelerates at a
rate of 45.3 m~s$^{-2}$ for about two hours before leveling out at a
final speed of 439 km~s$^{-1}$.

\begin{figure}[t]
\plotfiddle{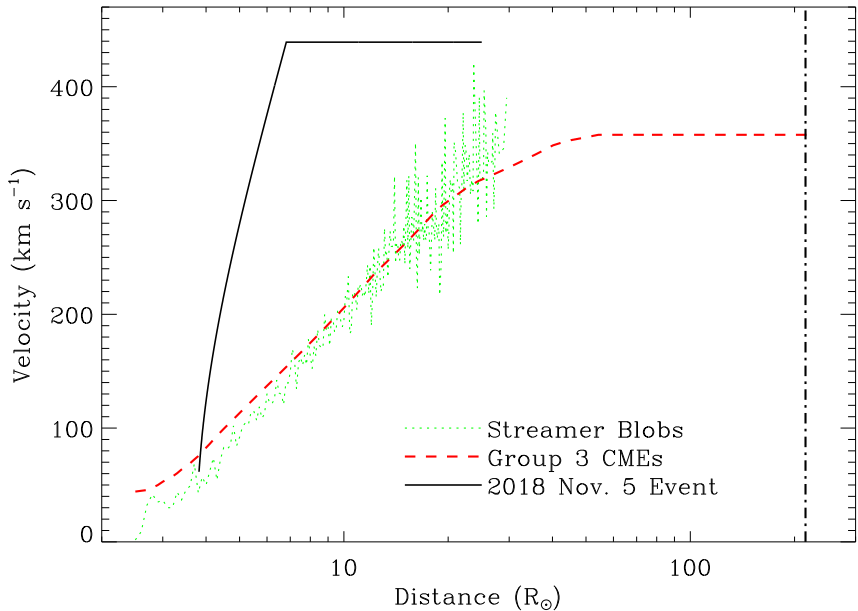}{2.8in}{0}{70}{70}{-220}{-265}
\caption{Velocity is plotted versus distance from Sun-center for
  the 2018~November~5 CME based on the kinematic model from
  Figure~7.  This is compared with an average streamer blob
  kinematic profile based on measurements from Wang et al.\ (2000),
  and an average kinematic profile of Group~3 CMEs from
  Wood et al.\ (2017), which are CMEs with no associated solar
  surface activity.  The dot-dashed line marks the 1~au distance.}
\end{figure}
     Although the final speed of the CME is not particularly fast, and
is comparable to slow solar wind speeds, the kinematic profile still
involves a surprisingly impulsive acceleration.  In order to illustrate
this, in Figure~8 we compare the CME's kinematic behavior with that
typically observed for streamer blobs and slow CMEs.
The streamer blob profile is based on velocity-vs.-distance
measurements of about 80 events from \citet{ymw00}.  We place
these measurements into 0.1~R$_{\odot}$ distance bins, and compute the
average velocity within each bin, yielding the blob kinematic profile
in Figure~8.  For the slow CMEs, we compute the average kinematic
profile of the ``Group~3'' CMEs in the \citet{bew17} sample of
Earth-directed events, where ``Group~3'' CMEs are ones with no
accompanying surface activity (e.g., flares or filament eruptions).
These CMEs overlap the ``streamer blowout'' category of transients
\citep{rah85,av17}.  At this point, we
should mention that inspection of images from the {\em Solar Dynamics
Observatory} reveals no evidence of surface activity
associated with the November~5 CME, but its trajectory would suggest a
source $17^{\circ}$ behind the limb as viewed from Earth, so
we cannot be completely certain that no surface activity accompanies
the eruption.

     The streamer blobs and Group~3 CMEs have essentially identical
kinematic profiles in Figure~8, involving a very slow acceleration
that does not reach a terminal velocity until $\sim 30$ R$_{\odot}$.
This behavior is widely assumed to be similar to that of the slow
solar wind.  In contrast, the 2018~November~5 transient reaches its
peak speed of 439 km~s$^{-1}$ by $\sim 7$ R$_{\odot}$.  The
transient is very much like previously studied streamer blobs
in general appearance, and
is clearly associated with the quiescent streamer structure.
However, its kinematics are very unusual for streamer blobs,
which were first defined by \citet{nrs97}.
More recent surveys of the blobs also do not include any events
that are so impulsively accelerated \citep{hqs09,clp18}.
The only clear exceptions are
blob-like ejections that follow CMEs \citep{hqs12}, but
there is no CME precursor for the November~5 event.

\section{Summary}

     We have studied a small transient observed by {\em PSP}/WISPR
on 2018~November~5, which is also seen in LASCO and COR2-A
coronagraphic images from {\em SOHO} and {\em STEREO-A}, respectively.
Our findings are summarized as follows:
\begin{description}
\item[1.] Despite looking narrow and jet-like from 1~au, the
  WISPR images are very suggestive of an FR morphology,
  with a visible leg stretching back towards the Sun long after
  the leading edge of the transient has left the FOV.
\item[2.] Assuming an FR shape, we perform a 3-D reconstruction
  of the small CME, representing the first such reconstruction
  considering both multiple 1~au vantage points and a viewpoint
  close to the Sun.  Synthetic images from the reconstruction
  are reasonably successful at reproducing the CME appearance
  in the WISPR, LASCO, and COR2-A images.
\item[3.] A kinematic fit to measurements from LASCO images
  implies that the CME accelerates at a rate of about 45.3 m~s$^{-2}$
  to a terminal speed of 439 km~s$^{-1}$, which it reaches at a
  distance of about 7~R$_{\odot}$ from Sun-center.
\item[4.] The small transient looks very much like a streamer blob in
  the coronagraph images from 1~au, but its kinematics are unusual, with a
  somewhat higher speed, and a much faster acceleration.  A more
  extensive observational study is necessary to determine whether the
  2018~November~5 transient is representative
  of a class of streamer disconnection events that are clearly
  distinct from the streamer blobs studied in the past, or if the
  event is best described as being simply a blob with anomalously
  impulsive acceleration.
\end{description}


\acknowledgments

Financial support was provided by the Chief of Naval Research, and by
NASA award 80HQTR18T0084 to the Naval Research Laboratory.

\end{document}